\documentclass[seceq]{ptptex}

\usepackage{graphicx}
\usepackage{wrapft}
\usepackage{fancybox}
\usepackage{subfigure}

\newcommand{\beq}{\begin{eqnarray}}
\newcommand{\eeq}{\end{eqnarray}}
\newcommand{\bfl}{\begin{flushleft}}
\newcommand{\efl}{\end{flushleft}}

\notypesetlogo                       
\title{
$\kappa$ Particle in the Analysis of the BESII Data and Chiral $\sigma$ Nonet}

\author{
Kunio \textsc{Takamatsu}
}

\inst{
KEK, High Energy Accelerator Research Organization, \\
Tsukuba, Ibaraki 305-0801, Japan\\
{\rm (For the sigma group)}
}

\abst{
The observation of the scalar meson, $\kappa(900)$ was reported 
in the fall of the last year in the analyses on the 
$K\pi$ system of the $J/\psi \to \bar{K}^{*} (892)^{0} K^{+} \pi^{-}$ 
decay data obtained by the BES collaboration. 
The analyses have been performed by two groups of the collaboration. 
The results obtained in the both analyses are in good agreement. 
They are also in good agreement with those obtained in the analysis of 
$D^{+} \to K^{-}\pi^{+}\pi^{+} $ decay by the E791 experiment. 
The evidence for the existence of the $\kappa (900)$ particle 
in the production process as well as in the scattering process 
confirms the scalar $\sigma$ nonet with that of $\sigma(600)$, firmly. 
It may be the chiral scalar $\sigma$ nonet which is 
the chiral partner of the $\pi$ nonet. Here, the essential points of 
the analysis of $\kappa(900)$ in the BES $J/\psi$ decay data, 
a brief review on present status of the observation of $\sigma(600)$ and $\kappa(900)$ 
and also the chiral scalar $\sigma$ nonet are presented. 
It is also tried to summarize the relation between the $\pi\pi$ ($K\pi$) 
scattering and the production amplitudes, relating with comments which were 
presented in the course of studies of the $\sigma(600)$ and the $\kappa(900)$ particles. 
The comments concern with the apparent differences between the $\pi\pi$ ($K\pi$) 
scattering phase shift data and the mass spectra of the $\pi\pi$ ($K\pi$) system 
in the production processes observed experimentally.
}
\begin{document}
\maketitle
\section{Introduction}
The evidence of the $\kappa(900)$ has been observed\cite{ref1} in the analysis of the $K\pi$ 
scattering phase shift data\cite{ref2} and 
in the production process, 
$D^{+}\to K^{-} \pi^{+} \pi^{+}$ decay by the E791 experiment at Fermilab.\cite{ref3} 
Recently, it was also observed in the analyses\cite{ref4} on the $K\pi$ system produced 
in the $J/\psi \to \bar{K}^{*}(892)^{0}K^{+}\pi^{-}$ decay data obtained with 
BESII at BEPC. The $\kappa(900)$ particle may be considered to be an 
ingredient of the scalar $\sigma$ nonet with $\sigma(600)$\cite{ref5}. 
It is different from the $SU(3)$ $^{3}P_{0}$ nonet, and may be the chiral 
partner of the $\pi$ nonet\cite{ref6,ref7}.
The analyses in the BES collaboration have been performed 
by two groups
{\footnote{
Wu Ning group (IHEP) and the sigma group (KEK, Nihon University and 
other universities: Shin-ichi Kurokawa, Kunio Takamatsu, Tsuneaki Tsuru, 
Kumatarou Ukai (KEK), Masuho Oda (Kokushikan U), Muneyuki Ishida (Meisei U), 
Tatsurou Matsuda, Yuya Toi (Miyazaki U), Shin Ishida, Toshihiko Komada, 
Tomohito Maeda, Kenji Yamada (Nihon U), Ichirou Yamauchi (TMCT)). 
The sigma group joined the BES collaboration in 2000, aiming at studies 
of chiral particles in the analyses on $J/\psi$ decays data obtained by BESII.
}} 
with different PWA methods. 
The results obtained in the both analyses are consistent well, 
determining the resonance parameters in good accuracy. 
They are also consistent well with those of E791. 
It is vitally important for the studies of low mass scalars
 that $\kappa(900)$ has been observed in the production process 
as well as in the scattering process. It may be worthwhile to be 
mentioned here that the studies with the BES $J/\psi$ decay data 
started early of 2000 and the preliminary results\cite{ref8,ref9} 
in the analyses were presented at the early stage of the analyses. 
\begin{figure}[htbp]
\begin{tabular}{cc}
\begin{minipage}{0.47\hsize}
\begin{center}
\includegraphics[width=\hsize]{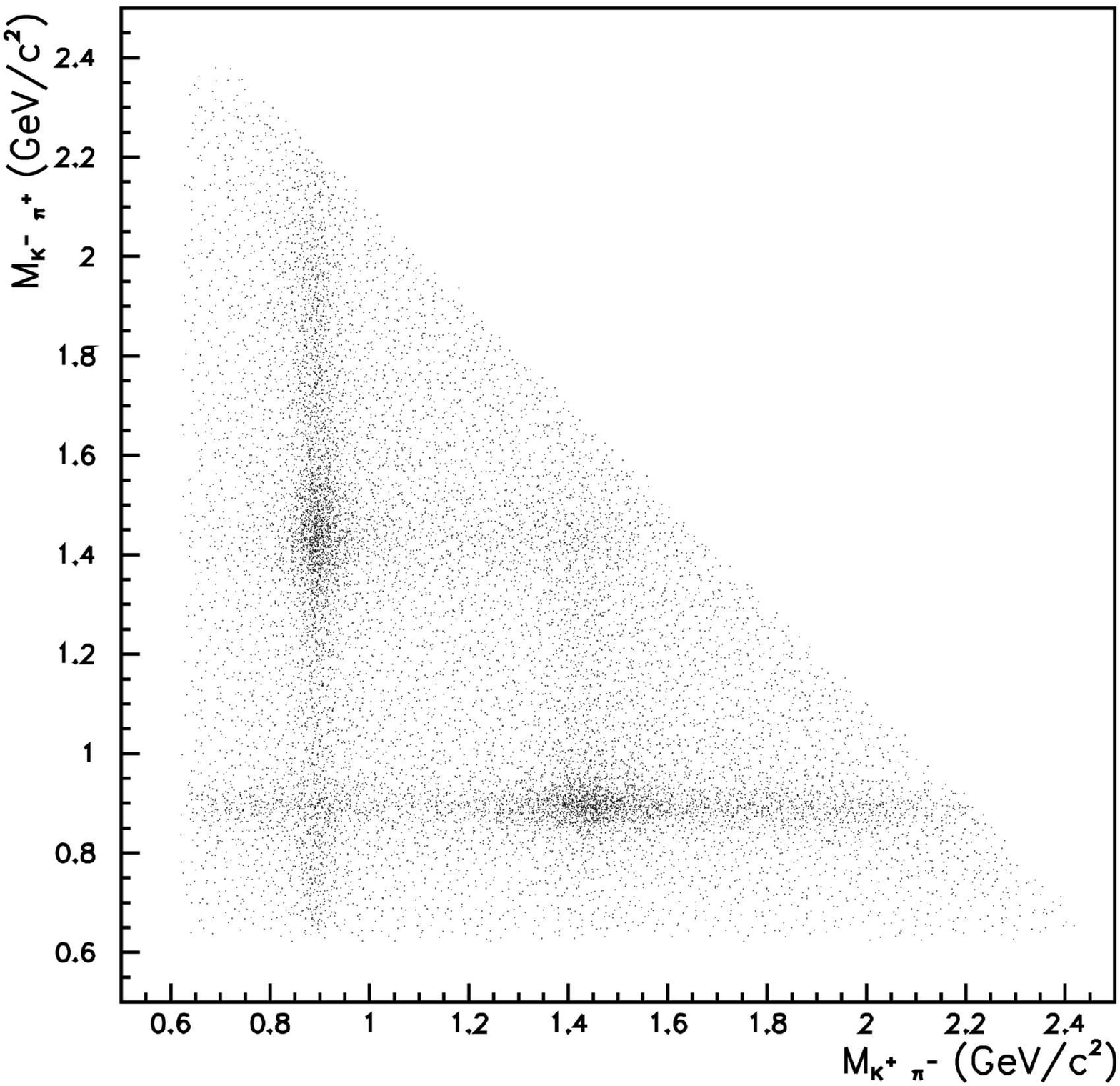}
\caption{Scatter plot of $M_{K^{+}\pi^{-}}$ versus $M_{K^{-}\pi^{+}}$ \protect\\
for $J/\psi\to K^{+}K^{-}\pi^{+} \pi^{-}$ of the BESII data.}
\label{fig:1}
\end{center}
\end{minipage}
\begin{minipage}{0.47\hsize}
\begin{center}
\includegraphics[width=\hsize]{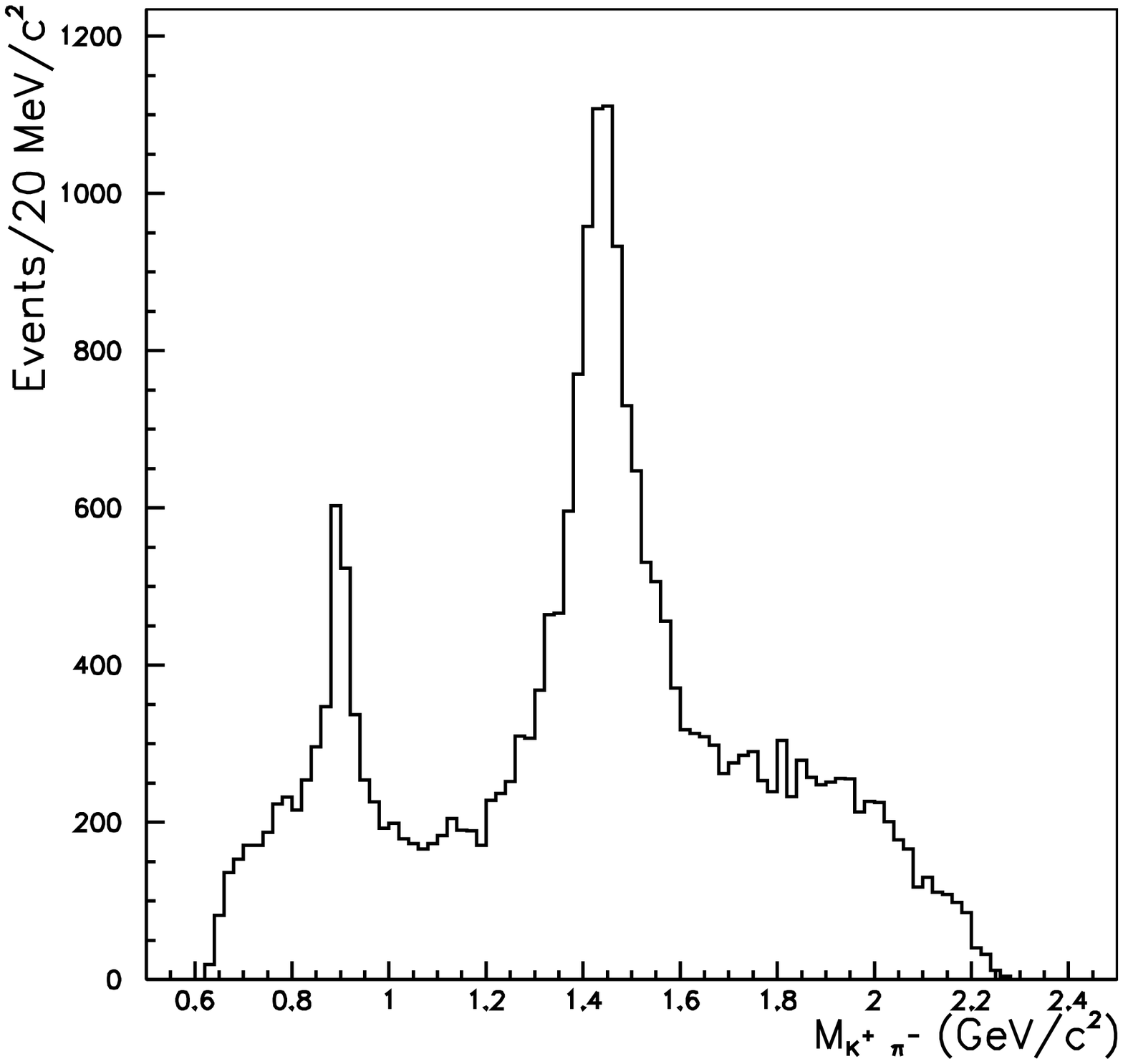}
\caption{Invariant mass spectrum for $K^{+}\pi^{-}$ associated with $K^{*}(892)^{0}$.}
\label{fig:2}
\end{center}
\end{minipage}
\end{tabular}
\end{figure}

The sigma group had obtained definite results of the $\kappa$ particle 
in the close collaboration with the Wu Ning's group in the analysis 
of the $J/\psi \to \bar{K}^{*}(892)^{0}K^{+}\pi^{-}$ decay. 
While the selection of the data was performed with the azimuth angle of 
$|\cos \theta | \leq 0.85$ at the early stage of our analysis, 
the final results were obtained in the analysis on 
the new data set selected with the narrower angle of 
$|\cos \theta | \leq 0.80$, based on the discussions of the 
collaboration at the later stage of the works. 
The SIMBES{\cite{ref10}} code was used for the detector simulation.

It was regrettable that the BES collaboration encountered, 
in the course of the analyses works on the $\kappa$ particle, 
with two troubles arising from a willful hope of a group in 
the BES collaboration who desired to report their $\kappa$ results 
in their analysis on a $K\pi$ system of four body state, $J/\psi\to KK\pi\pi$ in 
a certain journal combining with the results $\kappa$ 
in our analyses on the $K\pi$ system in the 
$J/\psi \to \bar{K}^{*}(892)^{0}K^{+}\pi^{-}$ decay data, 
though our results were almost at the final stage of the analyses. 
It was around the fall of 2002. 
Studies for $\kappa$ in PWA on the $KK\pi\pi$ channel has to treat a 
larger number of resonance states including sizable 
backgrounds more than those in PWA 
on the $K^{*}(892)K\pi$ channel. It is easy to see 
that the latter process is more suitable for studies 
to observe the $\kappa$ resonance than the former. 
The scatter plots of $K^{+}\pi^{-}$ versus $K^{-}\pi^{+}$ in fig. 1 
show the situation clearly, where most $K\pi$ events in the decay 
proceed with $K^{*}(892)$/ $\bar{K}^{*}(892)$. 
In fact, the $\kappa$ analysis on the $J/\psi\to KK\pi\pi$ decay data 
performed by a group was, however, not on $K\pi$ system in the 
$KK\pi\pi$ process but was on the $K\pi$ system recoiling 
against $K^{*}(892)$, 
as recognized in their articles presented at a later stage of their analysis. 
It was regrettable to see that their tricky and unfair request disturbed 
the BES collaboration and suffered its analysis works.
Moreover, the group who intended to analyze the $KK\pi\pi$ channel
insisted strongly to apply their analysis way to others. 
Their way for analysis has no physics base to be applied in the analysis, 
as will be discussed below in this report. The discussions concerning the 
analysis channels and the analysis methods brought confusions into the 
collaboration resulting in retardation of the studies. 
The collaboration decided finally that two reports of results 
in the analyses of $K^{*}(892)K\pi$ and those of $KK\pi\pi$ 
should be prepared independently each other and should be submitted to 
journal(s) separately at the same time{
\footnote{
While, a review article was reported in a 
journal which includes the results of $\kappa$ 
in the PWA analysis on the $K^{*}(892) K\pi$ system. 
A paper reporting their PWA results was also 
submitted to a journal. Both articles were 
not approved for publication by the collaboration. 
The latter report was withdrawn by receiving 
the complaint of the collaboration. 
}}. The one of the authors of the latter, nevertheless, 
published an article by a single name on PWA results of $\kappa$ particle 
in the spring of 2005. The collaboration had not recognized the fact, 
before they presented their statement at the plenary session 
of the Hadron conference, Hadron05 in Rio de Janeiro. 
It declared that the one is not the member of the collaboration 
any more and has no right to use the BES data. 
The whole members of the collaboration were surprised 
with one's violent act, extraordinary. 

The results of the $\kappa$ particle in the analyses of the $K\pi$ system 
of the $J/\psi \to \bar{K}^{*}(892)^{0}K^{+}\pi^{-}$ decay data were published 
in the fall of 2005, after the long time taken in the collaboration. 
It may be noted that the sigma group has obtained the solid recognition 
on the relation between the scattering and production amplitudes in 
the course of studies on the $\sigma$ particles which started ten years ago. 
The recognition relates with the cancellation mechanism\cite{ref11,ref12,ref13} 
in the scattering amplitudes. The one presented, however, 
discussions based on an understanding of the relation between 
the scattering and production processes, overlooking the cancellation mechanism. 
The discussions may be referred in one's related article\cite{ref14}. 
The one introduces an artificial suppression factor as an ``Adler zero''
which has no relation with the Adler self-consistency condition, 
and then, tries to perform combined 
fit of the scattering and production data and to deduce a phase motion 
of the production process which is obtained to be quite similar with 
that of the scattering process. 
These will be discussed with our criticisms as well as with the relation 
between the scattering and production amplitudes in the section 4. 
Before it, there will be described the essential points of 
the analysis of the $\kappa$ in the $J/\psi\to \bar{K}^{*}(892)^{0} K^{+}\pi^{-}$ 
decay data in the section 2 and of the analyses results of the $\sigma$ particle 
in the scattering and production processes and those of the $\kappa$ particle 
in the scattering process in the section 3. 
The chiral $\sigma$ nonet will briefly be described as well.

\section{The $\kappa$ particle in the analyses of the 
$J/\psi \to \bar{K}^{*}(892)^{0}K^{+}\pi^{-}$ decay of the BESII data}

The resonance parameters of the $\kappa$ particle 
have been obtained in the PWA analyses\cite{ref4} performed on the 
$J/\psi \to \bar{K}^{*}(892)^{0}K^{+}\pi^{-}$ decay of the BESII data 
with two methods, 
the method A and the method B of Wu Ning and of the sigma groups, respectively.

The channel, $K^{*}(892)\kappa$ decay shows relatively large branching ratio, 
as well as $K^{*}(892) K_{0}^{*}(1430)$ and $K^{*}(892) K_{2}^{*}(1430)$. 
The $K^{*}(892) K\pi$ decay channel is adequate to be studied for 
observation of the $\kappa$ particle. 
fig. 2 shows the invariant mass distribution of 
the $K^{+}\pi^{-}$ system recoiling against $\bar{K}^{*}(892)^{0}$. 
A $K^{*}(892)^{0}$ peak is recognized on a broad accumulation of 
events around $900 {\rm  \ MeV}/c^{2}$. 
It is a peak coming from the $K^{+}\pi^{-}$ events associated with 
$K^{-}\pi^{+}$ events which are occasionally in the region of 
$\bar{K}^{*}(892)$. 
A $J/\psi\to \bar{K}^{*}(892)^{0} K^{*}(892)^{0}$ decay is a 
process of SU(3) suppression. 
No isolated peak of $K^{*}(1410)^{0}$ is seen in fig. 2 due to the same suppression.
Other backgrounds contributing to the lower mass region are $K_{S}$ 
and a phase space production of 
$\bar{K}^{*}(892)^{0}K^{+}\pi^{-}$. 
One of the decay $\pi$ from $K_{S}$ in 
the $J/\psi\to \bar{K}^{*}(892)^{0} K_{S}$ decay is misidentified as $K$. 
These backgrounds are treated as non interfering processes.

\hspace{1em}Contributions from decays of $K^{*}(892)\pi$ of 
$J/\psi \to K K_{1}(1270/1400)\to$ $K$ \ $K^{*}(892)^{0}\pi$ are treated as 
coherent processes. The decay, $J/\psi\to b_{1}\pi$, $b_{1}\to K^{*}
K$ and the direct decay, $J/\psi\to \bar{K}^{*}(892)^{0} K^{+} \pi^{-}$ 
are also included as coherent processes. 
Contributions from other decay modes, $K\rho$ and 
$K^{*} (1430)$ from the $K_{1} (1270)$ decay are carefully studied, 
since it has relatively large decay branching ratios to $K\rho$ and $K^{*}(1430)\pi$. 
Their effects to the $\kappa$ parameters have been confirmed to be 
negligibly small. The interference effect in the cross region of the 
two $K^{*}(892)^{0}$ bands which are seen in the scatter plot of 
fig. 1 is carefully studied by Monte Carlo simulation, taking 
the detector acceptance and the width of $K^{*} (892)^{0}$ into account. 
The effects have been confirmed to be negligible. 

Here, the points of the analysis performed by the sigma group (with Method B) 
will be described, compactly. The PWA analysis has been performed 
by the variant mass and width method (the VMW method) \cite{ref11,ref17}, 
a covariant field theoretical approach with chiral symmetry describing 
a production process of the strong interaction. It is consistent with 
unitarity condition. The strong interaction is a residual interaction of QCD 
among color singlet bound states, $\phi_{i}$ of quarks, anti-quarks and gluons. 
Unstable particles are denoted by $\phi_{i}$ fields, as well as stable particles. 
The strong interaction Hamiltonian, $H_{\rm strong} (\phi_{i})$ 
describes the generalized S-matrix. The bases of generalized S-matrix 
are given by the configuration space of the multi-$\phi_{i}$ states. 
A relevant process is described by a coherent sum of many body decays 
including isobars and non resonant decays in the process. 
$H_{\rm strong}$ induces the various final state interactions 
reduced to the strong phases of the corresponding amplitudes. 
A propagator, $1/(m_{s}^{2}-s-i\epsilon)$ is replaced by 
$1/[m_{s}^{2}-s-i\sqrt{s} \Gamma(s)]$ in order to describe a unstable particle 
when the strong interaction acts on it. A production amplitude is, 
then, given as follows for the case of the $J/\psi\to K^{*}(892)K\pi$ decay; 
\beq
{\cal F}_{K^{*}(892)K\pi} &=& 
{\cal F}_{K^{*} K_{0}^{*}(1430)}+
{\cal F}_{K^{*} K_{2}^{*}(1430)}+\cdots 
+{\cal F}_{{\rm direct} K^{*}(892) K\pi}\nonumber\\
{\cal F}_{K^{*}\kappa}&=& r_{\kappa} e^{i\theta_{\kappa}} 
[(m_{\kappa} \Gamma_{\kappa})/(m_{\kappa}^{2}-s-i\sqrt{s} \Gamma (s))].
\eeq
Here, $r_{\kappa}$ is a coupling constant and 
$\theta_{\kappa}$ is the re-scattering phase 
between $K^{*}(892)$ and $\kappa$ due to the final state interaction. 
The last term is the Breit-Wigner form. Other processes for resonance states 
are described in the same manner.
All amplitudes concerning the process are considered: 
$a) \ J/\psi\to K^{*}(892)R_{K\pi}$, 
$b) \ J/\psi\to K R_{K^{*}(892)\pi}$, 
$c) \ J/\psi\to \pi R_{K^{*}(892)K}$, and 
$d) \ J/\psi \to K^{*}(892)K\pi$.  
$R_{K\pi}$ and $R_{K^{*}(892)\pi}$ stand for resonance states 
which decay to $K\pi$ and $K^{*}(892)\pi$, respectively. 
The process (c) is a direct decay of $J/\psi$ interfering with 
other resonance processes. S- and D-waves are considered for the 
intermediate $K\pi$ states, which have 
$J^{P}=0^{+}$ ($\kappa$, $K_{0}^{*}(1430)$) and $2^{+}$ 
($K_{2}^{*}(1430)$, $K_{2}^{*}(1922)$). 
The simplest Lagrangians relevant for production and decay of 
respective intermediate resonances are considered. 
In the case for the mechanism (b), intermediate $K^{*}(892)\pi$ resonances 
with $J^{P}=1^{+}$, $K_{1}(1270)$ and $K_{1}(1400)$ decaying 
into S-wave $K^{*}(892)\pi$ systems are acceptable. 
$b_{1}(1235)$ is considered for the $K^{*}K$ decay of the mechanism (c). 
The Lagrangians and amplitudes for the S-wave decay of the mechanism (a) 
(intermediate $K\pi$ resonances) are as follows; 
\beq
{\cal L}_S & \sim & 
\xi_{\kappa} \psi_\mu K^*_\mu \kappa + g_{\kappa} \kappa K \pi + \cdots,   \nonumber \\
{\cal F}_S & = & S_{h_\psi h_{K^*}}
(   r_{\kappa} e^{i \theta_{\kappa}} \Delta_{\kappa}(s_{K \pi})
 + r_{K_0^*} e^{i \theta_{K_0^*}}  \Delta_{K_0^*}(s_{K \pi})
+ r_{K \pi} e^{i \theta_{K \pi}} 
  ),   \nonumber \\
  &  &  \Delta_{\kappa}(s_{K \pi})  = \frac{ m_{\kappa} \Gamma_{\kappa}  }
{  m_{\kappa}^2 - s_{K \pi} - i \sqrt{s_{K \pi}} ~ \Gamma_{\kappa} (s_{K \pi}) }  
 ~~, \nonumber \\
 &  &  \Gamma_{\kappa} (s_{K \pi}) =  \frac{{\bf p} g_{\kappa}^2}{8 \pi s_{K \pi}}
\ \ .
\label{eq:2.1a}
\eeq
The factors $S$ is due to the helicity combination among relevant particles. 
The other processes for mechanisms, D-wave of (a) and (b), (c) and (d) 
are described elsewhere\cite{ref9}. The backgrounds are described above already for 
$K^{*}(892) \to  K\pi$, $K_{S} \to  \pi\pi$, one of the $\pi$ 
being misidentified to $K$, $K^{*}(1410)$ 
and $K^{*} K \pi $ phase space which are treated as non interfering processes. 
Then, the total amplitude F squared for the process is given by 
\beq
|{\cal F}|^{2} &\sim & 
|{\cal F}_{S}+{\cal F}_{D}+{\cal F}_{K_{1}} +
{\cal F}_{b_{1}} + {\cal F}_{\rm direct} |^{2} \nonumber \\
&&+(\sum |{\cal F}_{K^{*}P}|^{2}+|{\cal F}_{K_{S}}|^{2}
+|{\cal F}_{K^{*}(892)K\pi PS}|^{2}),
\eeq
where $K^{*}P$ stands for $K^{*}(892)$ or $K^{*}(1410)$. 

The $\bar{K}^{*}(892)^{0} K^{+} \pi^{-}$ events are selected with the requirement 
$0.812 {\rm  \ GeV}/c^{2} $$<$
$M_{K^{-}\pi^{+}}$ $<$0.972$ {\rm  \ GeV}/c^{2}$ 
for $\bar{K}^{*}(892)^{0}$ of the $J/\psi \to K^{+}K^{-} \pi^{+}\pi^{-}$ 
decay sample selected by the SIMBES code on the 
$58$ million $J/\psi$ decay data obtained by BESII at BEPC. 
The mass distributions of $K^{+}\pi^{-}$ and $\bar{K}^{*}(892)^{0}\pi^{-}$ and 
the angular distributions of $K^{+}$ below $1 {\rm  \ GeV}/c^{2}$ 
of the $K^{+}\pi^{-}$ 
mass and those above $1 {\rm  \ GeV}/c^{2}$ are submitted for the ${\chi}^{2}$ fitting, 
simultaneously. fig. 3 illustrates the results of the fitting obtained 
by the sigma group (Method B). The resonance parameters for $\kappa$ obtained 
in the fitting are as follows; 
\beq
M_{\kappa} = 896\pm 54^{+64}_{-44} {\rm  \ MeV}/c^{2} \ \ \ {\rm and} \ \ 
\Gamma_{\kappa} = 463\pm 88^{+55}_{-31} {\rm  \ MeV}/c^{2}. \nonumber
\eeq	
\begin{figure}[h]
 \begin{center}
  \begin{tabular}[t]{c}
   \subfigure[$K \pi$ mass spectrum ]
    {\label{subfig:01}\includegraphics[height=6cm,width=6cm,angle=0]{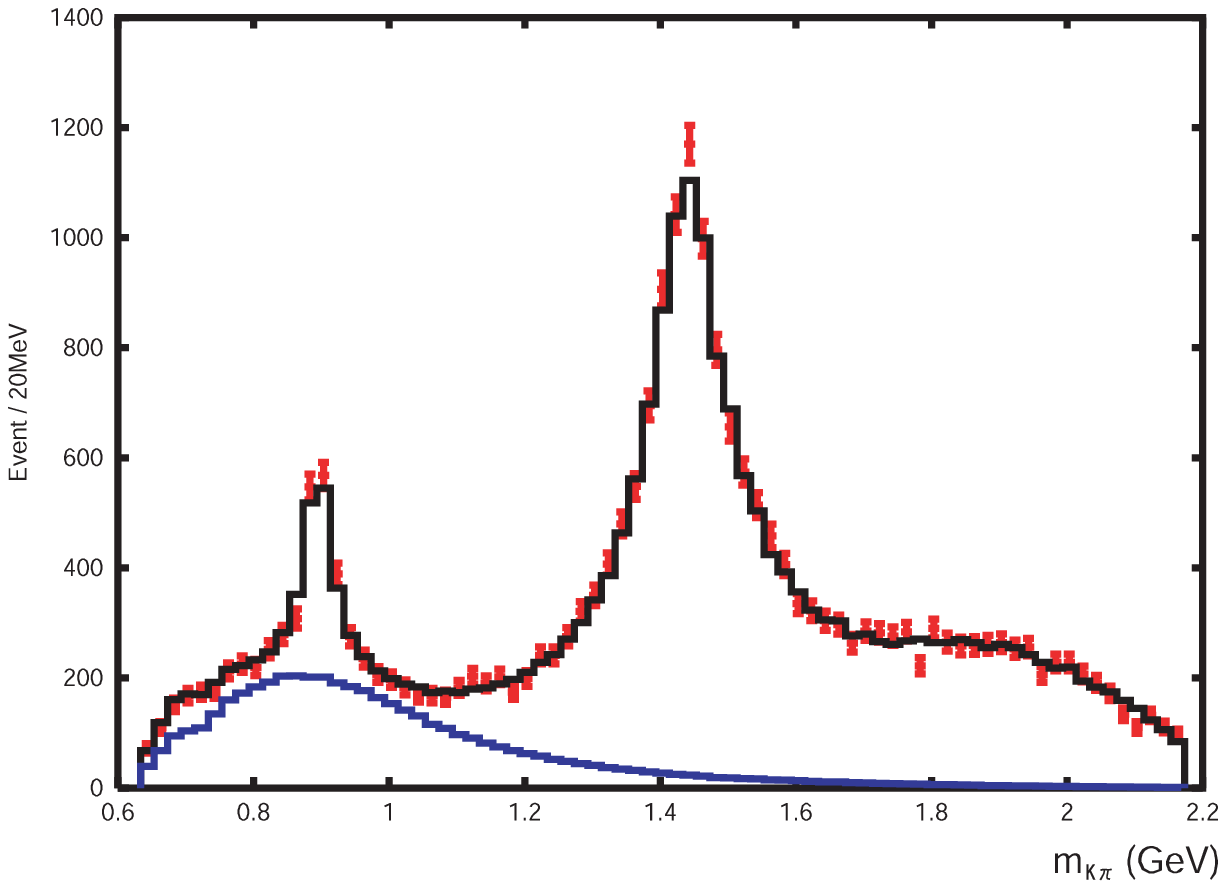}} 


   \subfigure[$K^* \pi$ mass spectrum ]
    {\label{subfig:02}\includegraphics[height=6cm,width=6cm,angle=0]{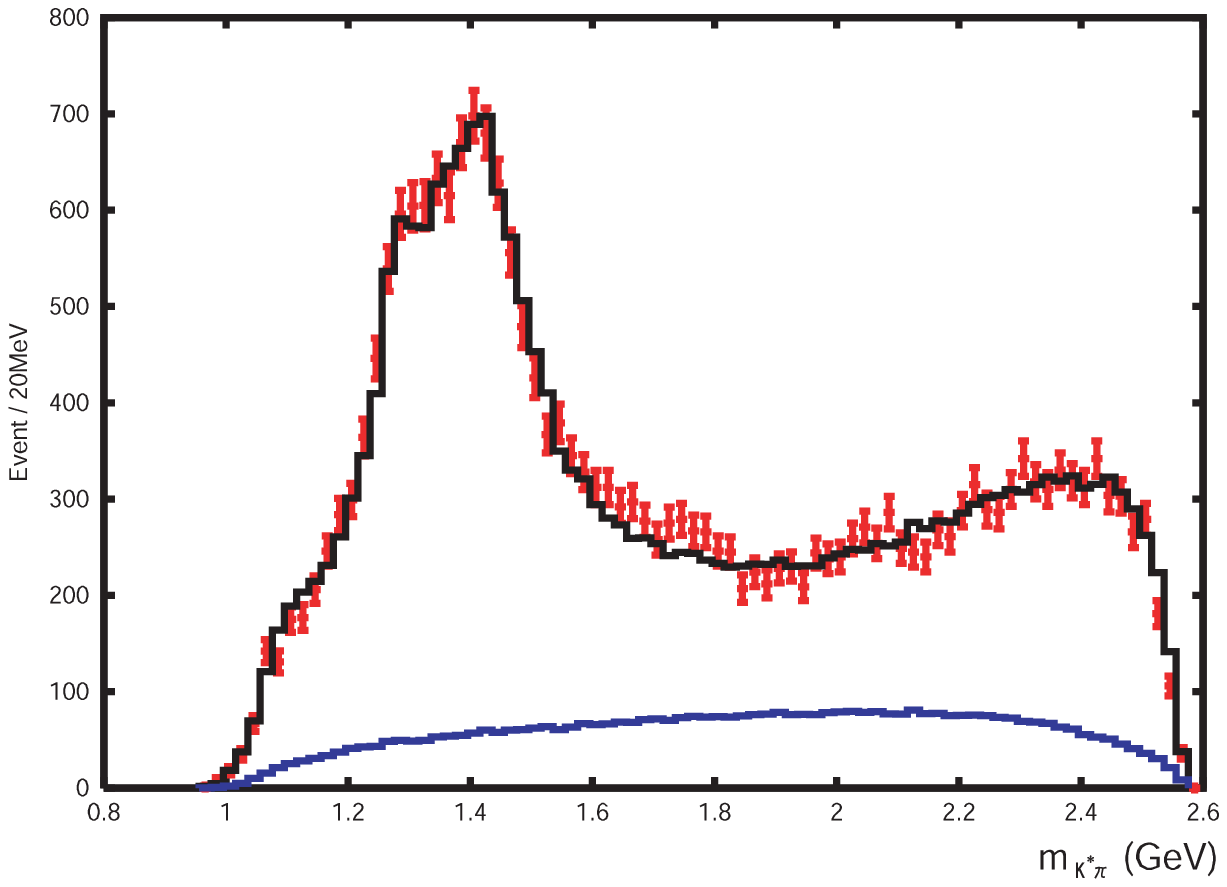}} 
    \vspace{-1em} \\

   \subfigure[$K^+$ angular distribution  $m_{K \pi} \leq 1$ (GeV)]
    {\label{subfig03}\includegraphics[height=6cm,width=6cm,angle=0]{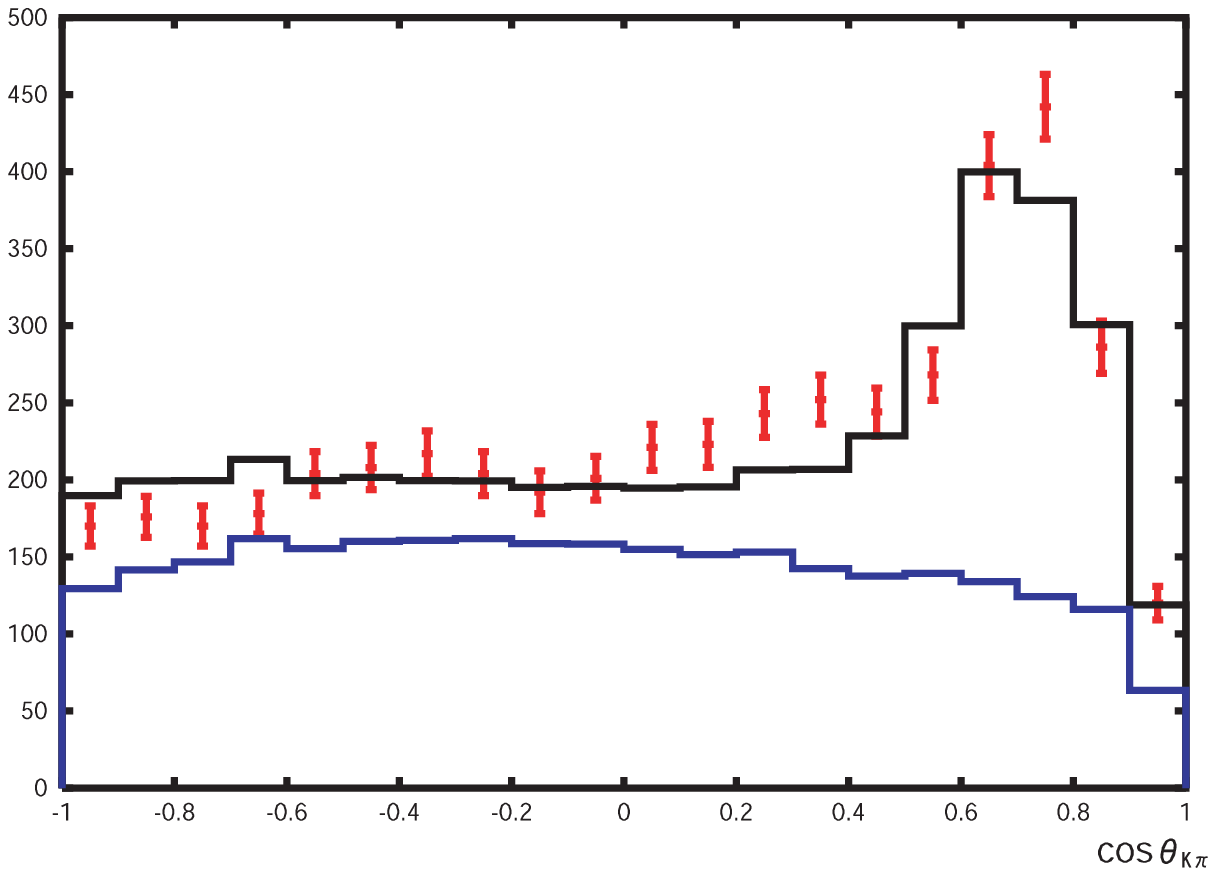}} 

   \subfigure[$K^+$ angular distribution $m_{K \pi} > 1 $  (GeV)]
    {\label{subfig:04}\includegraphics[height=6cm,width=6cm,angle=0]{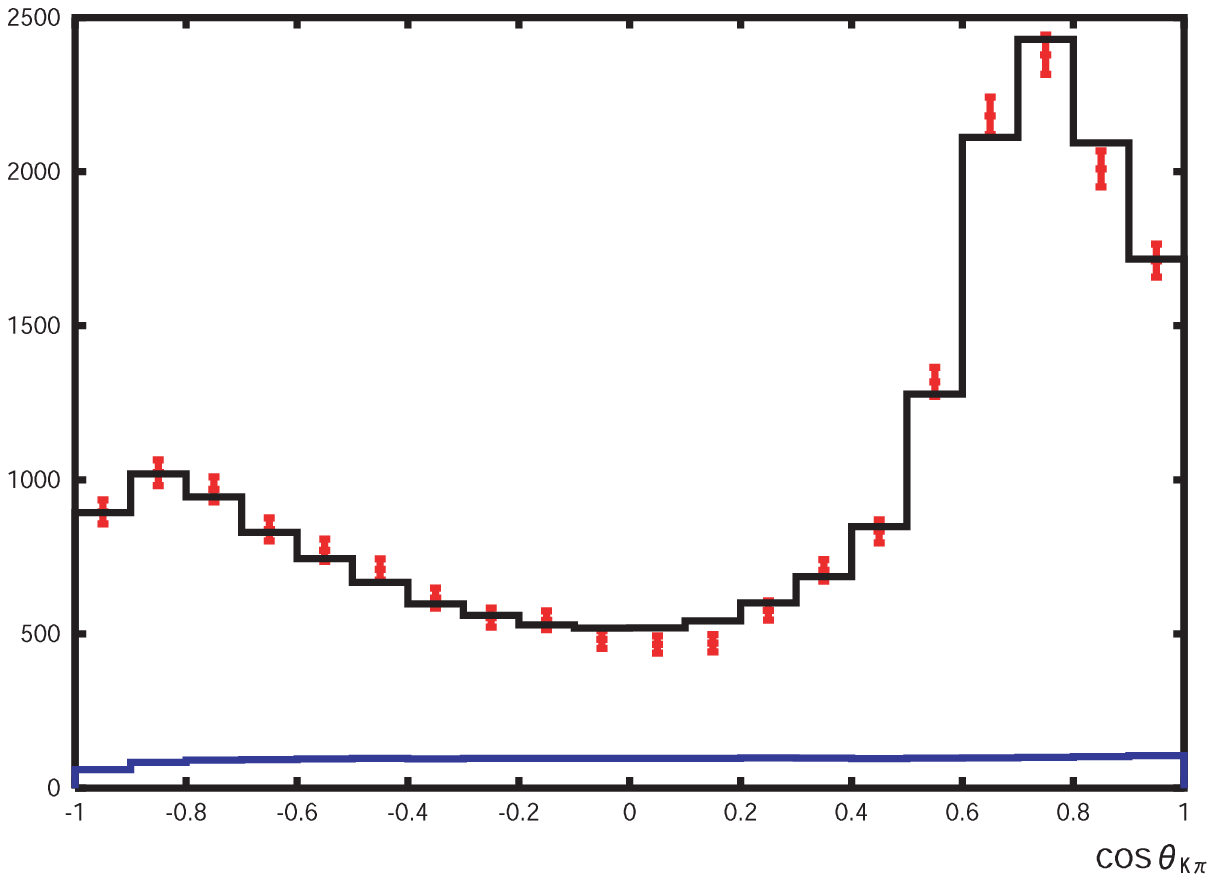}} 


  \end{tabular}
 \end{center}
\caption{Fitting results obtained by Method B. 
Solid points with error bars are the experimental data. 
Solid lines are fitting results. The contribution of $\kappa$ 
is illustrated in each figure. a) mass distribution of $K^{+}\pi^{-}$, 
b) mass distribution of $\bar{K}^{*}(892)^{0}\pi^{-}$, 
c) Angular distribution of $K^{+}$ (below $1 {\rm  \ GeV}/c^{2}$) 
and d) angular distribution of $K^{+}$ (above $1 {\rm  \ GeV}/c^{2}$).}
\label{fig:3}
\end{figure}
The errors are statistical (second term) and systematic (third and fourth terms). 
The difference between the results obtained by interfering and by 
non interfering amplitudes for $K^{*}(1410)$ and the effects of $1\sigma$ deviation 
for each resonance parameter obtained in the PWA are included 
in the evaluation of the systematic error. 
Above results agree well with those obtained by Method A, 
\beq
M_{\kappa} = 874 \pm 25^{+12}_{-55} {\rm  \ MeV}/c^{2} \ \ 
{\rm and} \ \ \ \Gamma_{\kappa} = 518\pm 65^{+27}_{-87} {\rm  \ MeV}/c^{2}.\nonumber
\eeq
They agree also well with those obtained in the re-analysis\cite{ref1} 
on the scattering phase shift data,
\beq
M_{\kappa} = 905^{+65}_{-30} {\rm  \ MeV}/{c}^{2} \ \ \ {\rm and} \ \ 
\Gamma_{\kappa} = 463^{+235}_{-110} {\rm  \ MeV}/c^{2}.\nonumber
\eeq
The averaged values are obtained for those of Method A and Method B,
\beq
M_{\kappa} = 878\pm 23^{+64}_{-55} {\rm  \ MeV}/c^{2} \ \ \ {\rm and} \ \ 
\Gamma_{\kappa} = 499\pm 52^{+55}_{-87} {\rm  \ MeV}/c^{2}.\nonumber
\eeq

The observation of $\sigma(600)$ and $\kappa(900)$ in production processes, 
as well as in the scattering process confirms the existence of the new scalar nonet, 
which might be assigned to be the chiral partner of the ground state $\sigma$ nonet. 
Efforts at the next stage of studies should be paid for searching low mass 
axial-vectors which may be ingredients of a new nonet, a chiral partner nonet 
of the ground state $\rho$ nonet. A scalar state with a negative charge 
conjugation parity will be another interesting object. 
The interesting level classification of the ground state meson nonets 
in the framework of the $\widetilde{U}(12)$ scheme was presented by Prof. Yamada\cite{ref15}, 
yesterday at the seminar. 
Though it was really the long way of the observation 
and the publication of the $\kappa(900)$ in the BES data, 
it is believed that they have opened the perspectives for studies 
of chiral particles at BES. 

There were cast, however, several comments 
relating analysis methods in the course of the present analyses 
on the $\kappa(900)$. 
They concerned all with the relation between the 
scattering and the production amplitudes. 
The criticisms on their comments will be described 
after brief description of the results of the analysis 
of the sigma group on $\sigma(600)$ in the production process 
and those on $\sigma(600)$ and $\kappa(900)$ 
in the scattering phase shift data in the next section. 
The chiral scalar $\sigma$ nonet will also be described briefly.

\section{Analyses of $\sigma(600)$ and the $\kappa(900)$ in the production 
and scattering processes and chiral scalar nonet}
\begin{flushleft}
{\it The low mass scalar, $\sigma$ in the central collision process}
\end{flushleft}
It took ten years from the observation of $\sigma(600)$ 
in the GAMS data to that of $\kappa(900)$ in the BES data. 
Studies of chiral particles were opened by 
the sigma group with the observation of the low mass scalar state 
in the $pp$ central collision production. The evidence of the $\sigma(600)$ 
in the reanalysis of the scattering phase shift data followed soon after it. 

The $\sigma(600)$ was observed\cite{ref16} in the studies of neutral mesons 
in the $\pi^{0}\pi^{0}$ state obtained by the GAMS4000 spectrometer at 
$450 {\rm  \ GeV}$ SPS at CERN. The $\pi^{0}\pi^{0}$ state was produced in the $pp$ 
central collision process, $pp \to p_{f}Xp_{s}$, $X \to {\pi}^{0} {\pi}^{0}$. 
The process is considered to proceed dominantly through 
pomeron-pomeron collision. The invariant mass spectrum showed a 
large event accumulation below $1 {\rm  \ GeV}/c^{2}$, 
which was analyzed by the variant mass and width (VMW) method\cite{ref11,ref17}. 
The result in the analysis is illustrated by a solid line in fig. 4. 
\begin{figure}[h]
\begin{center}
\includegraphics[width=0.6\hsize]{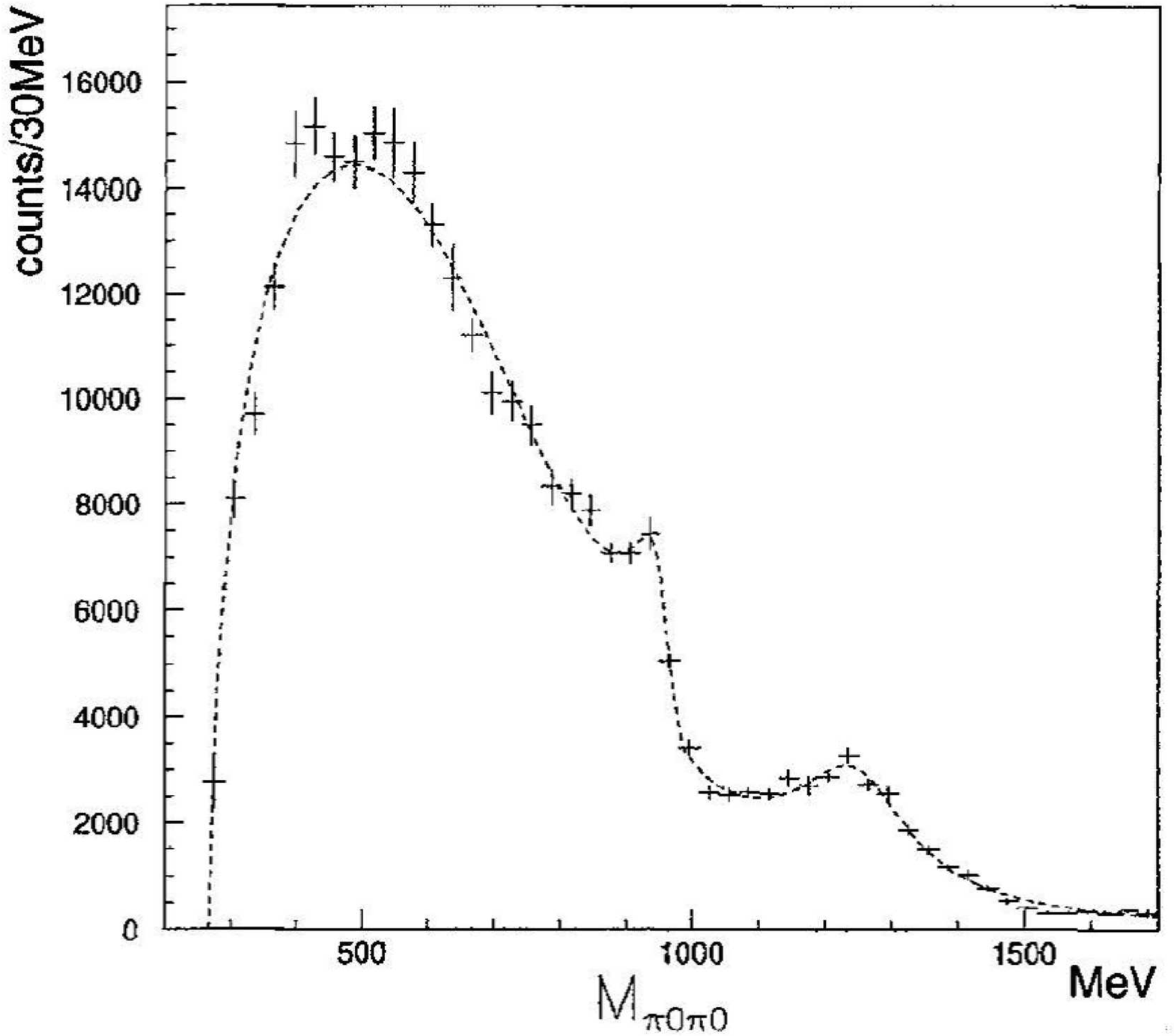}
\caption{The $\pi^{0}\pi^{0}$ 
invariant mass spectrum obtained 
in the pp central collision process. 
Solid line shows fitting in the analysis 
by the VMW method (by the GAMS collaboration).}
\label{fig:4}
\end{center}
\end{figure}
The resonance parameters are obtained as follows;
\beq
M_{\sigma} = 590 \pm 10 {\rm  \ MeV}/c^{2} \ \ \ {\rm and} \ \  
\Gamma_{\sigma} = 710\pm 30 {\rm  \ MeV}/c^{2}.\nonumber
\eeq
When the results in the production process were reported\cite{ref18}, 
comments\cite{ref19} were presented at Hadron05 that the 
observation breaks the unitarity condition based on the argument of $\pi\pi$ 
universality, since the $\sigma$ particle is not observed\cite{ref20} 
in the analysis of the $\pi\pi$ scattering phase shift data. 
It was already argued\cite{ref21} at the Hadron93 that the event 
accumulation below $1 {\rm  \ GeV}/c^{2}$ observed in the $\pi\pi$ 
spectrum obtained by the $pp$ central collision production at ISR at CERN 
was not a resonant signal. These comments were really intense at that time. 
The article\cite{ref20} was a sort of bible for the analysis on the $\pi\pi$ 
phase shift data. 
In fact, the low mass isoscalar scalar meson, $\sigma$ disappeared 
in the PDG tables from 1976 to 1996 for 20 years, 
till it revived in the tables with the rather strange name, 
$f_{0}(400-1200)/\sigma$.

The comments were presented in a definite way on two points. 
The $\sigma$ particle is not observed 
in the analyses on the scattering phase shift data. 
Then, the observation of the 
$\sigma$ particle in a production process breaks the unitarity condition. 
They concern with the recognition on the relation between
the scattering and the production amplitudes. 
The former was dependent on the recognition coming 
from overlooking the cancellation mechanism between $\sigma$ 
amplitude and non-resonant $\pi\pi$ amplitude which 
is guaranteed by chiral symmetry. 
The latter was dependent on elastic unitarity 
considering only $\pi\pi$ scattering{\footnote
{M.R. Pennington reported\cite{ref33} a result of rather 
large two photon coupling of the $\sigma$ resonance 
based on the recent results, $m_{\sigma} = 441^{+16}_{-8} {\rm  \ MeV}/c^{2}$, 
$\Gamma_{\sigma}= 544^{+18}_{-25} {\rm  \ MeV}/c^{2}$ obtained by Leutwyler et al.\cite{ref34}. 
Reanalysis of the $\pi\pi$ phase shifts data. 
}}. These are discussed below. 

The $\pi\pi$ phase shift data increase, as are well known, 
gradually from threshold to around $1 {\rm  \ GeV}/c^{2}$ 
by about $90$ degrees. Solid points in fig. 5 show the data. 
\begin{figure}[htbp]
\begin{center}
\includegraphics[width=0.5\hsize]{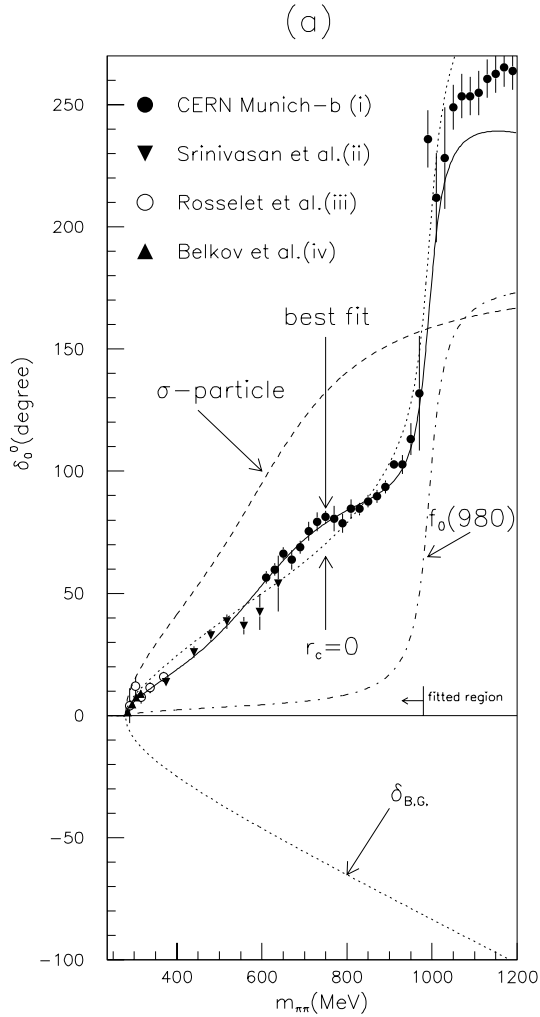}
\caption{$\pi\pi$ phase shift data (solid points) 
and the results in the analysis (solid and dotted lines). $\delta_{BG}$ 
is the phase shifts of the background due to the repulsive core. }
\label{fig:5}
\end{center}
\end{figure}
\begin{figure}[htbp]
\begin{tabular}{cc}
\begin{minipage}{0.45\hsize}
\begin{center}
\includegraphics[width=\hsize]{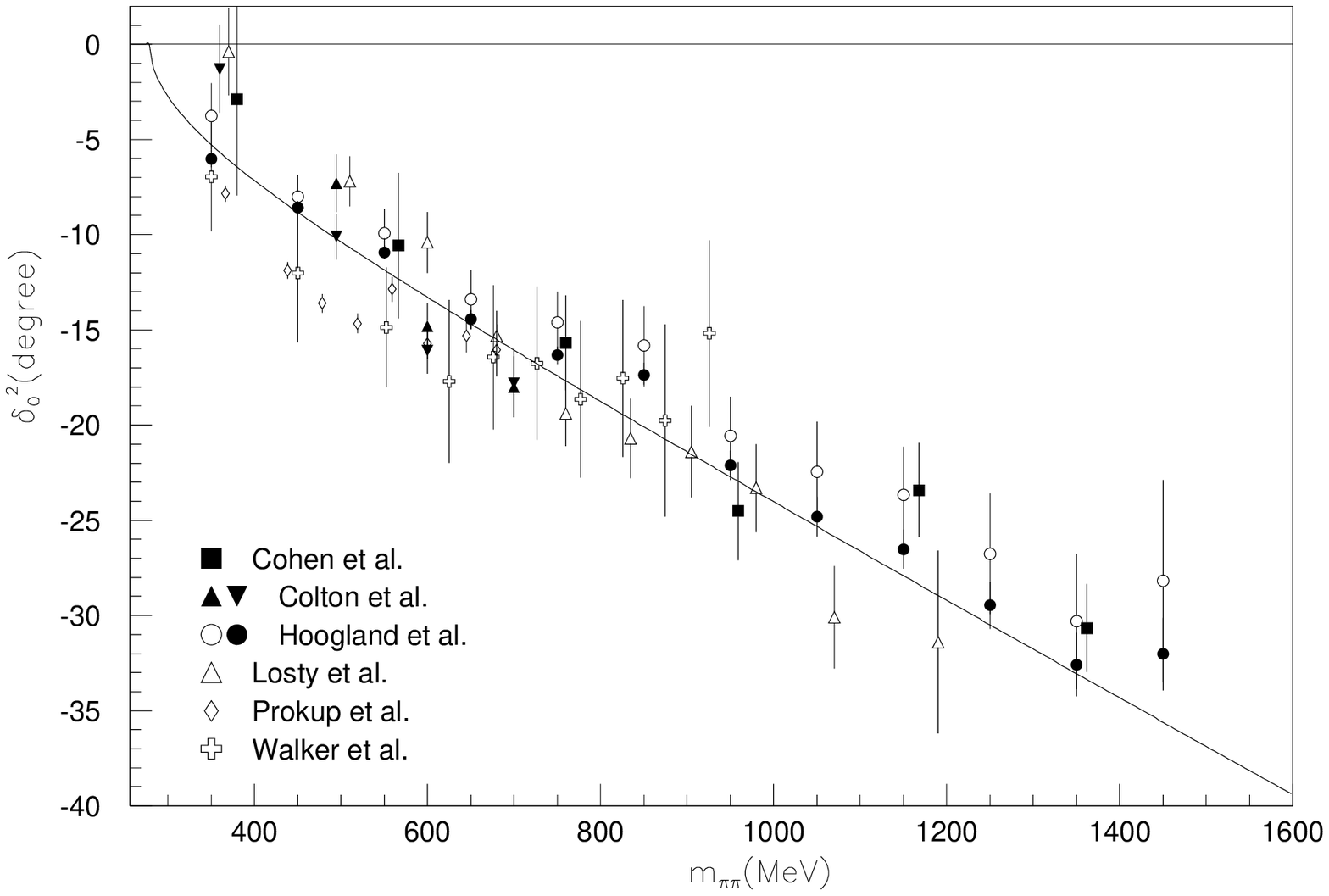}
\caption{$I=2$ phase shifts, $\delta_{0}^{(2)}$}
\label{fig:6}
\end{center}
\end{minipage}
\begin{minipage}{0.45\hsize}
\begin{center}
\includegraphics[width=\hsize]{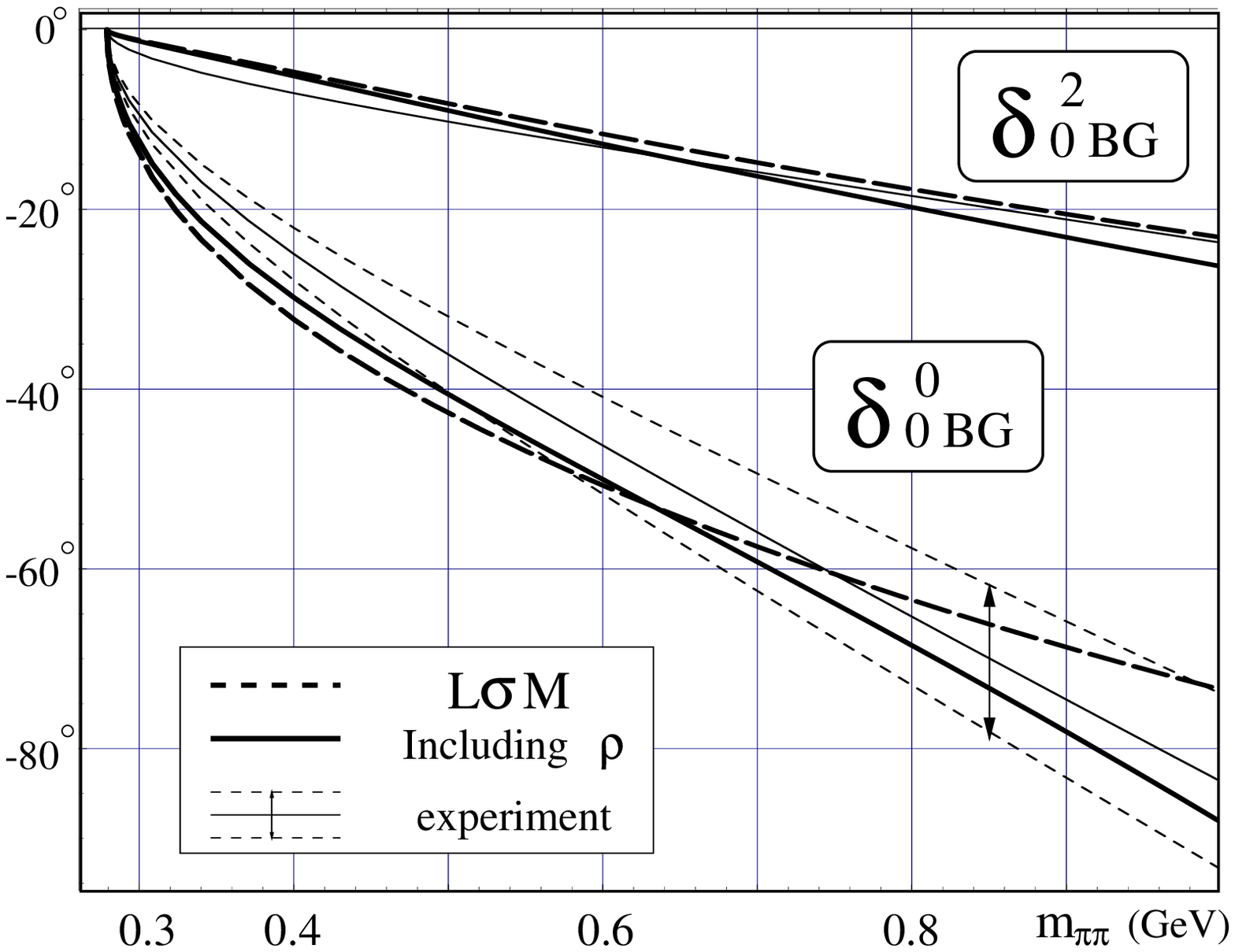}
\caption{$I=0$ background scalar phase shifts, $\delta_{0}^{0}{}_{BG}$ and $I=2$ 
background scalar phase shifts, $\delta_{0}^{(2)}$ estimated by the L$\sigma$M 
(Ref. 26)}
\label{fig:7}
\end{center}
\end{minipage}
\end{tabular}
\end{figure}
(The solid line and the dotted line show the results in the analysis 
performed by the sigma group, as are explained below.) 
One cannot recognize an enough phase variation for a resonance state in the region, 
if one considers only $\pi\pi$ interaction in an analysis. 
The events below $1 {\rm  \ GeV}/c^{2}$ are treated as a background, so far. 
Though the sizable events were recognized in data of production
 processes below $1 {\rm  \ GeV}/c^{2}$, they were 
forced to be treated as background.\cite{ref14, ref22} 

We reanalyzed\cite{ref23} the $\pi\pi$ phase shift data\cite{ref24}. 
The interference amplitude method (IA methods) was 
used in the analysis. It describes a process with physically 
meaningful parameters of resonances. Moreover, 
the background phase shifts coming from the 
repulsive core were considered properly and were 
introduced in the analysis. 

The S-matrix described as $S=e^{2i\delta (s)}=1+2i\alpha (s)$, where phase shift, 
$\delta (s)$ is a sum of phase shift of the resonance, 
$\sigma$ and that 
of the background due from the hard core below $K\bar{K}$ threshold, 
$\delta (s)= \delta_{R} +\delta_{BG}= \delta_{f_{0}}+\delta_{\sigma}+\delta_{BG}$ 
for the present case. 
$f_{0}$ and $\sigma$ are the relevant resonance states. 
The relativistic Breit-Wigner form is used for 
$\alpha (s)$, $\alpha (s)=s\Gamma_{R}(s)/[m_{R}^{2}-s-is\Gamma_{R} (s)]$. 
Accordingly, the total S-matrix is the product of 
the S-matrices of resonances, $S=S_{R} S_{BG}=S_{f_{0}} S_{\sigma} S_{BG}$. 
The unitarity condition is satisfied 
by unitarity of each S-matrix. 
A hard core type is used for the negative phase shifts, 
$\delta_{BG}=-r_{c}|\mbox{\boldmath $p$}_{1}|$, where $\mbox{\boldmath $p$}_{1}$ 
is momentum of $\pi$ 
in the center of mass system.

It is not an ad hoc idea to introduce the negative phase 
shifts but has the physics bases both experimentally and theoretically. 
The existence of negative phase shifts is observed in those of $I=2$ state, 
where no resonance state is expected, as shown in fig. 6. 
The phase shift decrease monotonically in proportion to the $\pi\pi$ mass 
from threshold to over $1 {\rm  \ GeV}/c^{2}$. 
\begin{table}[htbp]
\caption{Mass and width of $\sigma(600)$ obtained in analyses on 
$\pi\pi$ scattering and production processes}
\begin{center}
\begin{tabular}{lll}
\hline
\hline
\multicolumn{3}{l}{\bf $\sigma$ in scattering process (obtained by the Sigma group)}\\ 
S. Ishida et al.,
& $m_{\sigma} = 585\pm20 {\rm  \ MeV}/c^{2}$, 
& $\Gamma_{\sigma}=385\pm70 {\rm  \ MeV}/c^{2}$\\
\multicolumn{3}{l}{\bf $\sigma$ in production processes (obtained by the Sigma group)}\\
$pp$ central collision$^{\rm a)}$ 
& $m_{\sigma} = 590\pm10 {\rm  \ MeV}/c^{2}$,
& $\Gamma_{\sigma}= 710\pm30 {\rm  \ MeV}/c^{2}$\\
$J/\psi \to \omega\pi\pi$ (DM2 data)$^{\rm b)}$  
& $m_{\sigma} = 480\pm5 {\rm  \ MeV}/c^{2}$,
& $ \Gamma_{\sigma}= 325\pm10 {\rm  \ MeV}/c^{2}$\\
 $ p\bar{p} \to  3{\pi}^{0}$(CB data)$^{\rm c)}$
& $m_{\sigma} = 540^{+36}_{-29} {\rm  \ MeV}/c^{2}$,
& $\Gamma_{\sigma} = 385^{+64}_{-80} {\rm  \ MeV}/c^{2}$\\
$\Upsilon (3S)$, $\Upsilon (2S)$ decays$^{\rm d)}$ 
&$m_{\sigma} = 526^{+48}_{-37}{\rm  \ MeV}/c^{2}$,
&$\Gamma_{\sigma}= 301^{+145}_{-1000} {\rm  \ MeV}/c^{2}$\\
\multicolumn{3}{l}{\bf $\sigma$ in production processes} \\
$\tau$-decay (CLEO)$^{\rm e)}$ & $m_{\sigma} = 555 {\rm  \ MeV}/c^{2}$,
&$\Gamma_{\sigma} = 540 {\rm  \ MeV}/c^{2}$\\
$D \to   3\pi$ (E791)$^{\rm f)}$
&$m_{\sigma} = 483^{+25}_{-26}{\rm  \ MeV}/c^{2}$, 
&$\Gamma= 338^{+45}_{-42} {\rm  \ MeV}/c^{2}$\\
$D^{0} \to  K_{s}^{0} \pi\pi$ (CLEO)$^{\rm g)}$	
&$m_{\sigma} = 513 \pm 32 {\rm  \ MeV}/c^{2}$,
&$\Gamma_{\sigma}= 335\pm 67 {\rm  \ MeV}/c^{2}$\\
$J/\psi \to   \omega \pi\pi$ (BESII)$^{\rm h)}$
& $m_{\sigma}= 541 \pm 39 {\rm  \ MeV}/c^{2}$,
&$\Gamma= 504 \pm 84 {\rm  \ MeV}/c^{2}$\\
\hline
\end{tabular}
\end{center}
\footnotesize{
References\\
a) Ref. 12. 
b) K. Takamatsu et al., 
Proc. Int. Conf. Hadron Spectroscopy, BNL, Aug. 1997, AIP Conference 
Proceedings 432 (1997) p. 393. 
c) M. Ishida et al., Prog. Theor. Phys. \textbf{104} (2000) 203. 
d) T. Komada et al., Phys. Lett. B \textbf{508} (2001) 31; 
M. Ishida et al., Phys. Lett. B \textbf{518} (2001) 47. 
e) D.M. Asner et al., Phys. Rev. D \textbf{61} (2001)012002. 
f) E.M. Aitala et al., Phys. Rev. Lett. \textbf{86} (2001) 770 
g) H. Muramatsu et al., Phys. Rev. Lett. \textbf{89} (2002) 251802. 
h) M. Ablkim et al., Phys. Lett. B \textbf{598} (2004) 149.}
\end{table}
The existence of the repulsive core is also expected and deduced by the compensating 
$\lambda \phi^{4}$ contact interaction based on current algebra and PCAC. 
The $I=0$ and $I=2$ background scalar phase shifts, $\delta_{0}^{0}{}_{BG}$ and 
$\delta_{0}^{(2)}{}_{BG}$, 
respectively, are estimated by L$\sigma$M.\cite{ref26} 
The results obtained by M. Ishida are shown in fig. 7. 
The results of $\delta_{0}^{(2)}{}_{BG}$ are well consistent 
with those obtained in experiments shown in fig. 6. 

The hard core radius, $r_{c}^{0}$ due to $\delta_{0}^{0}{}_{BG}$ 
is treated as a fitting parameter in our analysis. 
The dotted line in fig. 5 illustrates the phase shifts, 
$\delta_{0}^{0}{}_{BG}$ obtained in the fitting. 
They are  well consistent with those estimated by L$\sigma$M in fig. 7. 
The result of the fitting on the overall $\pi\pi$ phase shifts 
is shown in fig. 5 by a solid line, which reproduces the data excellently. 
The resonance parameters, mass, width and core radius are obtained as follows;
\beq
M_{\sigma}&=&585 \pm 20 {\rm  \ MeV}/c^{2}, \ \ \ 
\Gamma_{\sigma}=385 \pm 70 {\rm  \ MeV}/c^{2} \nonumber\\ 
{\rm and} \ \ r_{c}^{0} &=&3.30 \pm 0.35 ({\rm  \ GeV}/c^{2})^{-1} (0.60 \pm0.07 {\rm fm}).
\nonumber
\eeq
A comment was presented in the Conference Summary of Hadron97\cite{ref27} 
on the above results. It mentioned only that the analysis 
was no more than the simple $\chi^{2}$ improvement with 
increasing number of fitting parameters by the 
introduction of a trivial background, 
but mentioned no word on the repulsive core nor its physics origin.

On the second point whether unitarity condition is satisfied in the analysis, 
it should be noticed that the VMW method treats unstable particles properly, 
as bound states of quarks, anti- quarks and gluons, as well as stable particles, 
and takes them as the base of the S-matrix. 
This fact corresponds to the observation of the $\sigma$ 
particle in the $\pi\pi$ scattering phase shifts. 
Generalized unitarity is hold in the analysis. 

There are tabulated in Table 1 
mass and width parameters obtained in the scattering and production processes 
by the sigma group, as well as those in the production processes by several groups. 
The mass and width values range in $400 {\rm  \ MeV}/c^{2} \sim 600 {\rm  \ MeV}/c^{2}$ 
and $300 {\rm  \ MeV}/c^{2} \sim 400 {\rm  \ MeV}/c^{2}$, respectively. 
The analysis of the $\kappa$ particle is reviewed in the scattering process, 
very briefly, in the next section.\\

\begin{flushleft}
{\it Observation of the $\kappa$ particle in the scattering phase shift data}
\end{flushleft}
The $K\pi$ phase shifts data\cite{ref2} were analyzed and the $\kappa$ 
particle has been observed in the analysis.\cite{ref1} The IA method is used in 
the analysis on the scattering phase shift data. 
The method is the same that used in the analysis on the $\sigma$ particle. 
The situation surrounding the studies of the $\kappa$ 
particle is quite similar with that of the $\sigma$ particle. 
The $I=1/2$ scalar phase shift data, $\delta_{0}^{1/2}$ 
are shown with solid circles in fig. 8a). 
There are also shown those of $I=3/2$, $\delta_{0}^{3/2}$ in fig. 8b) for reference, 
which exhibit a hard core like behavior up to around $1.2{\rm  \ GeV}/c^{2}$. 
The phase shifts, $\delta_{\kappa}$, $\delta_{K^{*}(1430)}$ and $\delta_{\rm BG}$ 
are considered for those contributing the process. 
The result obtained in the fitting for the phase shifts 
is illustrated by the solid line, which reproduced well the data. 
The dotted line shows the contribution from the $\kappa$ resonance. 
Be noted that $\delta^{1/2}_{\rm BG}$ is plotted with negative sign (positive value), 
since the narrow space for figure. 
The resonance parameters of the $\kappa$ particle, mass, width and hard core radius, 
$r_{c}^{1/2}$ obtained in the analysis are as follows; 
\beq
M_{\kappa}&=&905^{+65}_{-30} {\rm  \ MeV}/c^{2}, \ \ 
\Gamma_{\kappa} = 545^{+235}_{-110} {\rm  \ MeV}/c^{2} \nonumber\\ 
{\rm and} \ \  r_{c}^{1/2} &=& 3.57^{+0.4}_{-0.45} ({\rm  \ GeV}/c^{2})^{-1}
 (0.70^{+0.08}_{-0.09} {\rm fm}).\nonumber
\eeq
\begin{figure}[htbp]
\begin{tabular}{cc}
\begin{minipage}{0.47\hsize}
\begin{center}
\includegraphics[width=\hsize]{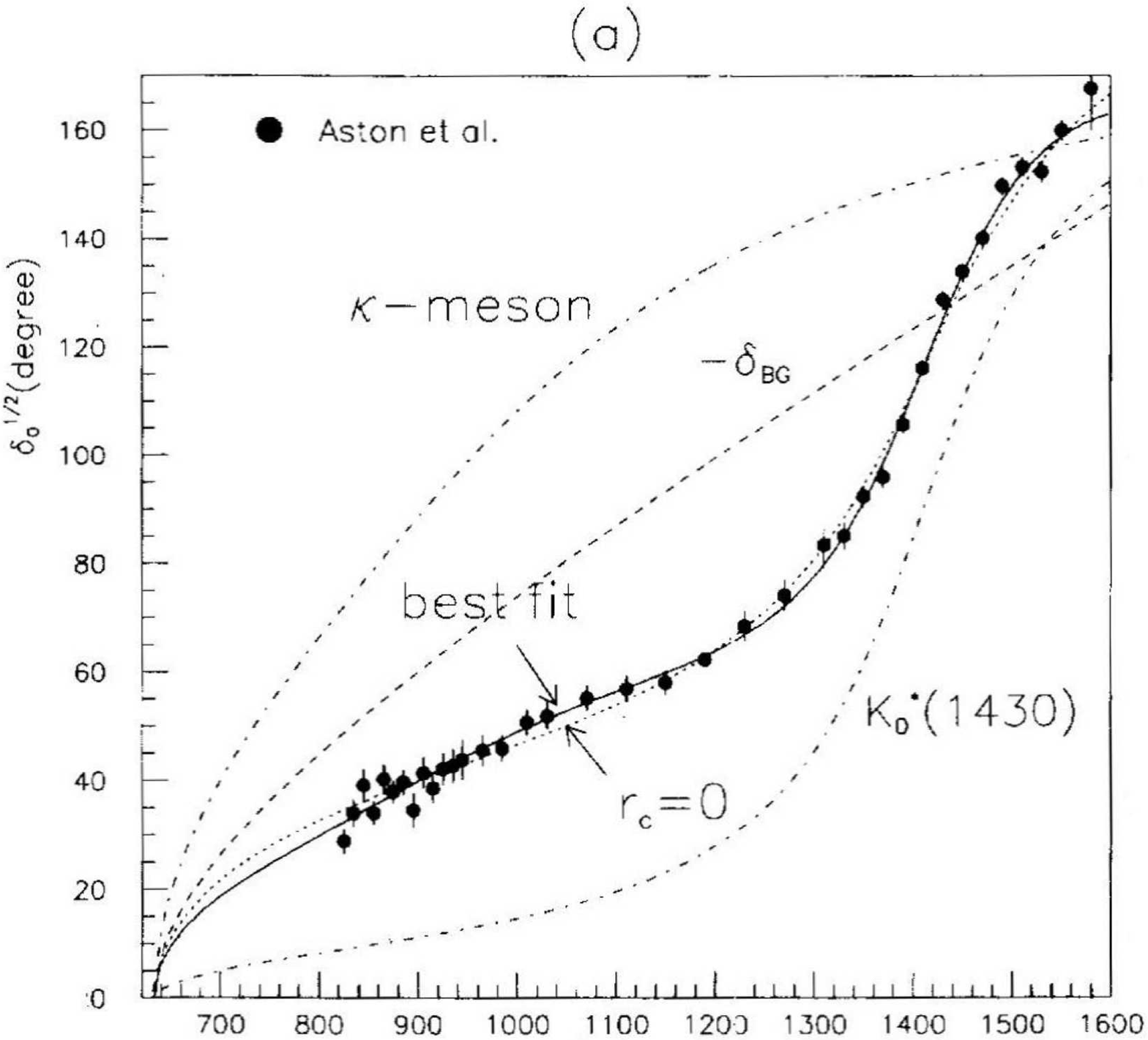}
\end{center}
\end{minipage}
\begin{minipage}{0.47\hsize}
\begin{center}
\includegraphics[width=\hsize]{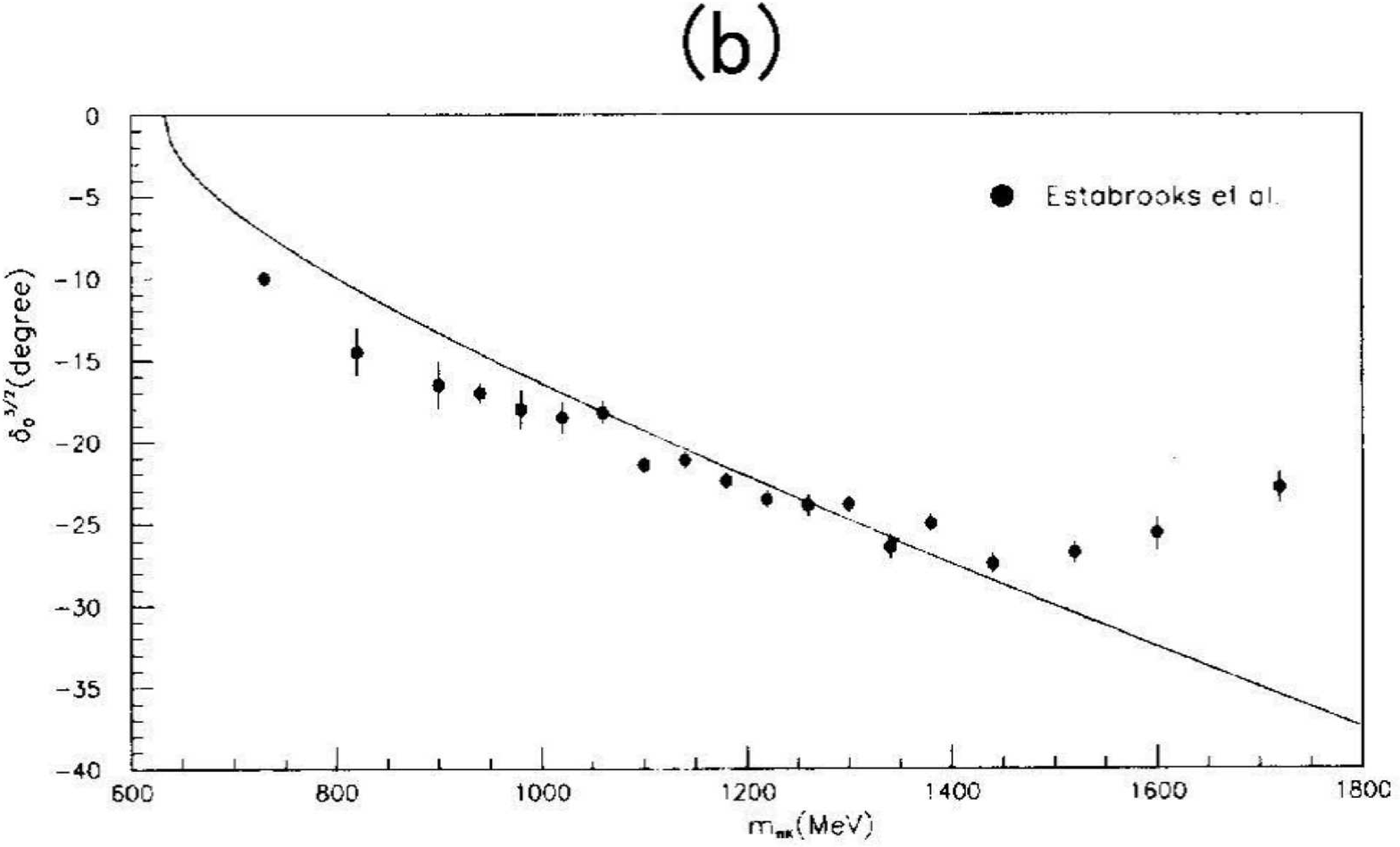}
\end{center}
\end{minipage}
\end{tabular}
\caption{a) Experimental data of $\delta_{0}^{1/2}$ (solid circles) 
and results of the analysis (solid line). $\delta_{0}^{1/2}{}_{\rm BG}$ 
is presented with negative sign (dotted line), 
since limited space for the figure. The case for $r_{c}^{0}=0$ 
is also sown (dotted line). b) $I=3/2$ phase shift, $\delta_{0}^{2/3}$ (Ref. 25) }
\label{fig:8}
\end{figure}
\begin{table}[htbp]
\caption{Mass and width of $\kappa(900)$ obtained in analyses on $K\pi$ 
scattering and production processes.}
\begin{center}
\begin{tabular}{lll}
\hline
\hline	
\multicolumn{3}{l}{\bf $K\pi$ scattering}\\
E. van Beveren et al.$^{\rm a)}$,&$m_{\kappa}=727 {\rm  \ MeV}/c^{2}$,
&$\Gamma_{\kappa} =526 {\rm  \ MeV}/c^{2}$\\
S. Ishida et al.$^{\rm b)}$,& $m_{\kappa}=905^{+63}_{-30} {\rm  \ MeV}/c^{2}$,
&$\Gamma_{\kappa}=545^{+235}_{-110} {\rm  \ MeV}/c^{2}$ \\ 
D. Black et al.$^{\rm c)}$, & $m_{\kappa}=897 {\rm  \ MeV}/c^{2}$,
&$\Gamma_{\kappa}=322 {\rm  \ MeV}/c^{2}$\\
J.A. Oller and E. Oset$^{\rm d)}$,&$m_{\kappa}=779 {\rm  \ MeV}/c^{2}$,
&$\Gamma_{\kappa}=660 {\rm  \ MeV}/c^{2}$\\
M. Jamin et al.$^{\rm e)}$,&$m_{\kappa}=708 {\rm  \ MeV}/c^{2}$,
&$\Gamma_{\kappa}=605 {\rm  \ MeV}/c^{2}$\\
D. Bugg$^{\rm f)}$,& $m_{\kappa}=722\pm 60 {\rm  \ MeV}/c^{2}$,
&$\Gamma_{\kappa}=772\pm 100 {\rm  \ MeV}/c^{2}$\\
H.Q. Zheng et al.$^{\rm g)}$, &$m_{\kappa}=594\pm 79 {\rm  \ MeV}/c^{2}$,
&$\Gamma_{\kappa}=724\pm 332 {\rm  \ MeV}/c^{2}$\\
F.K. Guo et al.$^{\rm h)}$,&$m_{\kappa}=757\pm 33 {\rm  \ MeV}/c^{2}$,
&$\Gamma_{\kappa}=558\pm 82{\rm  \ MeV}/c^{2}$\\
Z.Y. Zhou et al.$^{\rm i)}$,&$m_{\kappa}=694\pm 53 {\rm  \ MeV}/c^{2}$,
&$\Gamma_{\kappa}=606\pm 59{\rm  \ MeV}/c^{2}$\\
\multicolumn{3}{l}{\bf $K\pi$ production}\\
E.M. Aitala et al.$^{\rm j)}$,& $m_{\kappa}=797\pm 19\pm 43 {\rm  \ MeV}/c^{2}$,
&$\Gamma_{\kappa}=410\pm 43 \pm 87 {\rm  \ MeV}/c^{2}$\\
M. Ablikim et al.$^{\rm k)}$, & $m_{\kappa}=878\pm 23^{+64}_{-55} {\rm  \ MeV}/c^{2}$,
&$\Gamma_{\kappa}=499\pm 52^{+55}_{-87} {\rm  \ MeV}/c^{2}$\\
\hline
\end{tabular}
\end{center}
{\footnotesize 
References\\
a) Ref. 28. 
b) Ref. 1. 
c) D. Black et al., Phys. Rev. D \textbf{58} (1998) 054012. 
d) J.A. Oller and E. Oset, Phys. Rev. D \textbf{60} (1999) 074023. 
e) M. Jamin et al., Nucl. Phys. B \textbf{587} (2000) 331. 
f) ref. 14. 
g) H.Q. Zheng et al., Nucl. Phys. A \textbf{733} (2004) 235. 
h) Presented at the seminar: 
Results obtained in a re-analysis with chiral unitary approach. 
i) Presented at the seminar: Results obtained in the improved analysis 
of H.Q Zheng et al. j) Ref. 3). k) Ref. 4).}
\end{table}
In the case of no background due to the hard core, 
i.e. $r_{c}^{1/2}=0$, the mass value shifts at higher mass 
and shows an obscure resonance behavior with the worse $\chi^{2}$ value. 
Its result is also illustrated by the dotted line in the figure. 
The results show clearly the observation of the $\kappa$ particle 
in the scattering process.

Results in the analyses performed on the phase shift data by several 
groups are tabulated in the Table 2. 
Results obtained in the analyses on production processes are 
also summarized in the Table for those by E791\cite{ref3} on the $D$ decay
, $D^{+} \to K^{-}\pi^{+} \pi^{+}$ and by BES\cite{ref4} 
on the $J/\psi$ decay, $J/\psi \to \bar{K}^{*}(892)^{0}K^{+}\pi^{-}$
{\footnote{
The results in the analyses of the $\kappa$ particle on the 
$J/\psi \to \bar{K}^{*}(892) K^{+}{\pi}^{-}$ 
of the BESII data reported by D. Bugg are not included in the Table 2, 
since his work was not approved by the collaboration.
}}. 
The coherent S-wave contribution has been observed by FOCUS\cite{ref29} 
on the $D$ decay, $D^{+} \to K^{-} \pi^{+} \mu^{+} \nu$. 
However, observe no resonance state 
is observed in the phase motion by CLEO.\cite{ref30} 
A comment was presented for the existence of $\kappa$ 
similar to the case for that of the $\sigma$ particle.\cite{ref31}\\

\begin{flushleft}
{\it Chiral scalar $\sigma$ nonet}
\end{flushleft}

The observations of $\sigma(600)$ and $\kappa(900)$ 
in the scattering and production processes suggest strongly 
an existence of the chiral scalar nonet. 
It is different from the $SU(3)$ ${}^{3}P_{0}$ scalar nonet 
and considered to be the chiral partner of the ground state $\pi$ nonet\cite{ref6,ref7}. 
The mass and width parameters for $\delta / a_{0}(980)$ 
and $\sigma^{'}/f_{0}(980)$ are examined by M. Ishida\cite{ref7} 
based on the $SU(3)$ L$\sigma$M. 
The parameters are well consistent with experimental values, 
where $m_{\pi}$, $m_{\eta}$, $m_{\eta^{'}}$,$m_{s}$, $m_{\kappa}$ 
and $f_{p}$ are used as input parameters in the estimation. 
The chiral $\sigma$ nonet as the chiral partner of the ground state $\pi$ 
nonet can be identified to be 
$\{ \sigma(600), f_{0}(980), a_{0}(980), \kappa(900)\} $
{\footnote{A different assignment for $\kappa(900)$ 
is given by K. Yamada in the talk at the present seminar.} }.
E. van Beveren et al. suggested\cite{ref28} the $\sigma$ nonet 
by the unitarized coupled channel. 
It is interesting that Jaffe suggested\cite{ref32} also it by the 
$qq\bar{q}\bar{q}$ bag model. It will be the next task for 
the studies to search for a chiral axial-vector nonet, 
($f_{1}^{\chi}$,$f_{1}^{'\chi}$, $a_{1}^{\chi}$, $K_{1}^{\chi}$) 
as a chiral partner of the ground state $\rho$ nonet. 
It will also be interesting to search for other ground state chiralons, 
which may be a low mass extra scalar with $J^{PC}=0^{+-}$ 
and/or low mass vectors around 1.2  \ GeV, for example, 
in the level scheme of the $\widetilde{U}(12)$ classification. 
$J^{PC}=1^{-+}$ states which are assigned to be $L=1$ excited states 
in the $\widetilde{U} (12)$ classification are also very 
interesting to be searched for in the BESII data.
\section{Relation between scattering and production amplitudes and 
related, ``so-called Adler 0'',``combined fit'' and phase motion}
Rresults in an analysis of $\kappa$ 
performed by a group on the $K^{*}(892)K\pi$ channel in the BES 
$J/\psi$ decay data were reported in journals without an 
agreement of the BES collaboration as is mentioned in the introduction, . 
The acts violated the collaboration rule and 
were not acceptable by the collaboration. 
It is not a present aim to make a comment, 
here, on the results of its works, 
but it might be necessary to criticize, 
apart from the acts violating the collaboration rule, 
the analysis method described in the related article\cite{ref14} 
concerning with its analysis on the $\sigma$ and/or $\kappa$. 
The points to be criticized here is of ``so called Adler 0'' and of 
``combined fit on the data of scattering and production processes''. 
The points relate all with understanding on the relation between 
scattering and production amplitudes, 
which has been studied and clarified by S. Ishida and M. Ishida\cite{ref11,ref12} 
in the course of studies of $\sigma$ and $\kappa$. \\

\begin{flushleft}
{\it ``So-called Adler 0''}
\end{flushleft}

The $\pi\pi$ invariant mass spectra obtained experimentally 
in production processes appear apparently different from the 
scattering phase shift data, as were recognized in the processes 
of the $pp$ central collision, 
$J/\psi \to \omega \pi\pi$, $D\to 3\pi$, and others. 
The $K\pi$ system shows also different behaviors between 
the $K\pi$ scattering phase shifts data and the $K\pi$ invariant 
mass spectra of the production process, 
$J/\psi\to K^{*}K\pi$ of the BES experiment and $D\to K\pi\pi$ of E791. 
It has been mentioned in the previous section that 
both data in the scattering and production processes are not necessary 
to show the same behavior in each other, 
since there exist the cancellation mechanism 
in the scattering amplitudes. Nevertheless, 
the author of the articles proposes and insists to 
introduce an artificial suppression factor, $(s-s_{0})$ 
on a production amplitude to meet the scattering phase shift data, 
postulating that both scattering and production amplitudes should 
show the same behavior. 

It is believed generally that the production amplitude is 
connected with the scattering amplitude; 
${\cal F}=\alpha(s) {\cal T}$ and $\alpha (s)$ is 
the slowly varying real function. ${\cal F}$ has, 
consequently, the same phase that ${\cal T}$ has, and that the 
production process should be restricted by the unitarity condition 
based on the universality arguments for the $\pi\pi$($K\pi$) amplitudes 
of the scattering and the production processes. 
This is valid in an elastic unitarity case but is not applicable 
in the case where a resonance state(s) is exists. 
Despite, one proposes and insists to introduce an Adler 0 multiplying a factor 
$1/(s-s_{0})$ on $\alpha (s)$, in order to satisfy the conjecture, 
while the factor is artificial with no physics base. 
The suppression factor, $(s-s_{0})$ has no relation with 
the Adler self-consistency condition.\\

\begin{flushleft}
{\it ``Combined fit''}
\end{flushleft}

It is proposed to perform a so-called ``combined fit'' of both data 
in the scattering and production processes, simultaneously, 
based on introduction of an Adler 0. 
Consequently, the results of phase shifts of the production 
process become similar to the scattering phase shift data 
with rather a wide width value. 
The same analysis method including an Adler 0 is applied 
on the E791 data on $D^{+}\to \pi^{+} \pi^{+} \pi^{-}$, 
obtaining $494 {\rm  \ MeV}/c^{2}$\cite{ref14} for the width which 
increased about $50${\verb|%|} larger than the width value, $338 {\rm  \ MeV}/c^{2}$ 
reported by the E791 group
{\footnote{
The same treatment with an``Adler 0'' was also applied\cite{ref35} 
on the result of $D^{+}\to K\pi^{+}\pi^{-}$ obtained by E791\cite{ref3}, 
obtaining a wider width for the $\kappa$ resonance than the original one.}}. 
The mass value shifts to $533 {\rm  \ MeV}/c^{2}$ 
from the reported value $483 {\rm  \ MeV}/c^{2}$. 
Above values, $533 {\rm  \ MeV}/c^{2}$ and 
$494 {\rm  \ MeV}/c^{2}$ for mass and width, 
respectively, are quite similar to those, 
$525 {\rm  \ MeV}/c^{2}$ and 
$494 {\rm  \ MeV}/c^{2}$, respectively obtained in one's analysis 
on the $\pi\pi$ scattering data.\cite{ref14} 
These show that ``combined fit'' with an artificial suppression factor 
makes the resonance parameters of the production process 
reproduce those obtained in the analysis on the scattering phase shifts. 
These facts are recognized more clearly in the reduction of phase motion 
of the production amplitude. \\

\begin{flushleft}
{\it Observation of phase motion}
\end{flushleft}

It is tried in one's analysis to optimize a phase of the $\kappa$ 
amplitude in every sliced mass of $K\pi$. 
In the optimization of phase values, however, 
the rest of parameters of the $\kappa$ 
amplitude are fixed to those obtained in one's ``combined fit''. 
It may be easy to recognize in the combined fit that phase values 
obtained will reveal those of the scattering data. 
In fact, a phase motion thus obtained for the production process 
is quite similar to that of the scattering process. 

It is pity that there is no reference wave found with an 
adequate background phase in the $\kappa$ 
region of the $K^{*}(892)K\pi$ channel. 
The $K^{*}(1430)$ which interferes with $\kappa$ 
is of no use, since both amplitude and phase cannot 
be determined unambiguously with an amplitude composed 
of sum of two S-waves, even though the amplitude and the 
phase of $K^{*}(1430)$ are determined by Breit-Wigner parameters. 

The Breit-Wigner parameter for the width of the $\kappa$ resonance, 
$\sim 500 {\rm  \ MeV}/c^{2}$ in our analysis 
is consistent with that obtained in the analysis of the $D$ decay, 
$D^{+}\to K^{-} \pi^{+}\pi^{+} $by the E791 experiment. 
It is argued\cite{ref36} that the results of E791 are 
contradict with those in the analysis for 
the S-wave component of the process, $D^{+}\to K^{-}\pi^{+}\mu^{+}\nu $ 
obtained by the FOCUS experiment.\cite{ref29} \ 
The S-wave component has almost constant phase, $\delta =\pi /4 $ 
in the $\kappa$ mass region, $m_{\kappa}=0.8 - 1.0 {\rm  \ GeV}/c^{2}$ 
to be the same as the scattering phase shifts by LASS in the mass region, 
suggested by Minkowski and Ochs. 
The $K\pi$ system in the semi-leptonic $D$ decay, 
$D^{+}\to K^{-}\pi^{+}\mu^{+}\nu$, 
is isolated and can be expected to have 
a same behavior with the $K\pi$ system 
of the scattering process. 
On the other hand, the $K\pi$ system in the hadraonic $D$ decay, 
$D^{+}\to K^{-}\pi^{+}\pi^{+}$ is not isolated in the $K\pi\pi$ three bodies 
interfering each other. The production amplitudes are not necessary 
to appear the same as those of the scattering.    
\section{Concluding remarks}
The PWA results of the low mass scalar particle, $\kappa$ observed in the analyses by 
two methods, method A and method B on the $J/\psi \to K^{*}(892)K\pi$ decay data 
of BESII have been published in the journal. 
The Breit-Wigner parameters obtained agree well with those of the E791 experiment. 
$\sigma(600)$ and $\kappa(900)$ are considered to form the scalar 
$\sigma$ nonet with 
$f_{0}(980)$ and $a_{0}(980)$, 
\{$\sigma(600)$, $f_{0}(980)$, $a_{0}(980)$, $\kappa(900)$\}. 
It is different from the $SU(3)$ ground state ${}^{3}P_{0}$ scalar nonet and is 
considered to be the chiral scalar $\sigma$ nonet as the chiral partner 
of the ground state $\pi$ nonet. 
It has taken more than ten years after $\sigma(600)$ was observed in the 
production process by the sigma group of the GAMS 
collaboration in the analysis of the $\pi^{0}\pi^{0}$ state 
produced in the $pp$ central collision process at the $450 {\rm  \ GeV}$ SPS at CERN. 

The VMW method was used in the analysis of the $K\pi$ system of the 
$J/\psi\to K^{*}(892)K\pi$ decay by method B. It treats unstable particles, properly, 
which are the color singlet bound states of quarks, anti-quarks and gluons, 
as well as stable particles. It satisfies the unitarity condition 
based on chiral symmetry. It describes the process by the coherent sum 
of Breit-Wigner amplitudes. In the course of the 
analyses works on $\sigma(600)$ and $\kappa(900)$, the sigma group 
has obtained the clear recognition on the relation between 
the scattering and the production amplitudes. Based 
on the quark physics description of the process, 
the cancellation mechanism in the scattering process 
has been understood well. It is recognized that the resonance parameters 
in the production processes can be determined independently 
from those in the scattering process. 

In these 30 years, it was believed that the $\pi\pi$ elastic 
unitarity condition in the scattering process should 
hold in the production processes at the low energy 
based on the $\pi\pi$ universality. 
It is proposed and insisted to introduce an artificial factor, 
``so called Adler 0'' on a production amplitude and 
to perform so-called ``combined fit'' in order to meet 
the above mentioned belief. Their proposals might be 
based on misunderstanding of the cancellation mechanism 
in the scattering amplitudes. PWA parameters obtained in 
analyses of production processes based on the proposals become, accordingly, 
dependent on the scattering data and are distorted. 

Low mass axial-vectors, $a_{1}^{\chi}$, $f_{1}^{\chi}$ 
and others are the 
next objects to be searched for. They are expected 
to be chiral partners of the ground state $\rho$ nonet.  
An extra scalar with $J^{PC}=0^{+-}$ and low mass extra vector 
states around $1.2 {\rm  \ GeV}/c^{2}$ are also another state interesting 
to be studied. They will be chiral states in the level 
scheme of the $\widetilde{U}(12)$ classification.


\section*{Acknowledgements}

The author would express his thanks to Prof. S. Ishida for the discussions on 
the $\widetilde{U}(12)$ classification scheme with its underlying physics and also of studies 
on searching for chiral particles on the BES data. He would thank Prof. K. Yamada 
for discussions on the ground state chiral nonets based on the $\widetilde{U}(12)$ 
classification 
scheme and for discussions on searching for chiral particles in the BES $J/\psi$ 
decay data. 
He would also thank all the members of the sigma group for their cooperation throughout 
the analysis works. 
The discussions with and help of Dr. Wu Ning given to the sigma group are invaluable 
and are greatly appreciated for performing the analysis on the $\kappa$ 
particle on the BES 
$J/\psi$ decay data. He would thank Prof. Weiguo Li and Prof. Xiaoyan Shen who support 
the sigma group throughout the works. All the colleagues of the BES collaboration 
should be acknowledged for their supports for the works. 
The China-Japan collaboration project of JSPS (Contract No. JR-02-B4) is acknowledged 
for the support given to the sigma group to perform the present studies on chiral 
particles. The sigma group would also express his thanks for Prof. S. Kurokawa for 
his care on the project.

\end{document}